\documentclass{article}

\usepackage{microtype}
\usepackage{graphicx}
\usepackage{subfigure}
\usepackage{booktabs} 
\usepackage{amsmath}

\usepackage{hyperref}

\usepackage[accepted]{icml2020}

\icmltitlerunning{Enhancement of shock-capturing methods via machine learning}

\begin{document}

\twocolumn[
\icmltitle{Enhancement of shock-capturing methods via machine learning}
\icmlsetsymbol{equal}{*}

\begin{icmlauthorlist}
\icmlauthor{Ben Stevens}{caltech}
\icmlauthor{Tim Colonius}{caltech}
\end{icmlauthorlist}

\icmlaffiliation{caltech}{Department of Mechanical and Civil Engineering, California Institute of Technology, Pasadena, CA, United States}

\icmlcorrespondingauthor{Ben Stevens}{bstevens@caltech.edu}

\icmlkeywords{Machine Learning, ICML}

\vskip 0.3in
]




\begin{abstract}
In recent years, machine learning has been used to create data-driven solutions to problems for which an algorithmic solution is intractable, as well as fine-tuning existing algorithms. This research applies machine learning to the development of an improved finite-volume method for simulating PDEs with discontinuous solutions. Shock capturing methods make use of nonlinear switching functions that are not guaranteed to be optimal. Because data can be used to learn nonlinear relationships, we train a neural network to improve the results of a fifth-order WENO method. We post-process the outputs of the neural network to guarantee that the method is consistent. The training data consists of the exact mapping between cell averages and interpolated values for a set of integrable functions that represent waveforms we would expect to see while simulating a PDE. We demonstrate our method on linear advection of a discontinuous function, the inviscid Burgers' equation, and the 1-D Euler equations. For the latter, we examine the Shu-Osher model problem for turbulence-shockwave interactions. We find that our method outperforms WENO in simulations where the numerical solution becomes overly diffused due to numerical viscosity.
\end{abstract}

\section{Introduction}
For some initial-boundary value problems (IBVP) in fluid mechanics, the solution of the partial differential equations (PDE) include discontinuous initial data or a discontinuity that forms in finite time, i.e. shockwaves. Numerical methods for solving these PDE must be specially tailored to properly resolve these discontinuities \cite{leveque2002finite}. 

These shock-capturing methods are designed with the goal of sharply resolving a shock without inducing spurious oscillations, while also giving accurate solutions in smooth regions of the flow. One major breakthrough in this effort was the development of high-resolution methods \cite{harten1983high} , as these methods were capable of achieving second-order accuracy without introducing spurious oscillations around shocks. These methods gave rise to a class of high-resolution methods called essentially non-oscillatory (ENO) schemes \cite{harten1987uniformly} that measure the smoothness of the solution on several stencils, and then compute the flux based on the smoothest stencil to avoid interpolating through the discontinuity.  These schemes are nonlinear (even when the PDE are linear) since the interpolation coefficients depend on the solution. These ideas were then modified to create WENO-JS (weighted ENO-Jiang Shu) methods \cite{jiang1996efficient}, which again compute the smoothness on several stencils. However, instead of taking only the smoothest stencil, these methods take a weighted average of the fluxes predicted on each stencil to emphasize the smoother ones. When each stencil is equally smooth, the weights are designed to cause the method to converge to the constant coefficient scheme that maximizes the order of accuracy over the union of the sub-stencils, which gives these methods a high order of accuracy for smooth solutions. 

Many efforts have built on the original WENO-JS schemes by modifying the smoothness indicators \cite{ha2013improved, kim2016modified, rathan2018modified}, modifying the nonlinear weights \cite{borges2008improved, castro2011high, rathan2018improved}, and using WENO-JS as part of a hybrid scheme \cite{li2010hybrid, peer2009method}. While some of these references build off of each other rather than starting from WENO-JS, we will base our method on WENO-JS because our strategy for developing the method does not resemble other methods. However, our methodology could easily adopt various improvements that have been made to WENO-JS.

One commonality that has persisted since the original ENO scheme is a reliance on human intuition in shock-capturing method design, particularly in the nonlinear aspects of the schemes, i.e. smoothness indicators and weighting functions.  While they have been well studied, there is no reason to believe that they are optimal. Efforts have been made to develop optimal spatial discretization methods by minimizing wave propagation errors \cite{kim1996optimized, lele1992compact, liu2013globally, tam1993dispersion} and minimizing error over certain frequency ranges \cite{zhang2013optimized}, and some of these techniques have even been combined with shock-capturing schemes \cite{fang2013optimized, wang2001optimized}. However, designing the optimization problem still requires human intuition with regards to balancing competing goals, rather than attempting to learn an optimal scheme from data in an unbiased way. 

Over the past decades, machine learning has become ubiquitous in data analysis and is increasingly seen as having potential  to improve (or reformulate) numerical methods for PDEs. Lagaris et al. \cite{lagaris1998artificial} parameterized the solution to a PDE as a neural network and optimized the weights to minimize the residual of the solution. Yu et al. \cite{yu2018data} trained a neural network to classify the local smoothness and apply artificial viscosity based on this classification. Bar-Sinai et al. \cite{bar2019learning} used simulation data to embed coarse graining models into finite difference schemes involving neural networks, allowing them to achieve low error on relatively coarse grids. Pfau et al. \cite{pfau2018spectral} parameterized the eigenfunctions of eigenvalue problems as a neural network and cast the training as a bilevel optimization problem to reduce bias, resulting in significantly decreased memory requirements. Hsieh et al. \cite{hsieh2019learning} attempted to learn domain specific fast PDE solvers by learning how to iteratively improve the solution using a deep neural network, resulting in a 2-3 times speedup compared to state of the art solvers. 

In the current work, we attempt to train a neural network to improve WENO5-JS. Our goal is to get closer to the optimal nonlinear finite-volume coefficients while introducing a minimal amount of bias. Unlike other references, we do not directly change the smoothness indicators or nonlinear weights of the method. Instead, we use a neural network to perturb the finite-volume coefficients determined using the original smoothness indicators and nonlinear weights of WENO5-JS. We attempt to learn an optimal function for this perturbation using data generated from waveforms that are representative of solutions of PDEs. These modifications result in a finite-volume scheme that diffuses fine-scale flow features and discontinuities less severely than WENO5-JS. We start in the next section by giving a more detailed description of the proposed algorithm.

\section{Numerical Methods}
\subsection{Description of WENO-NN}
Although we focus on WENO5-JS in this paper, our approach could generally be used to enhance any shock capturing method (or perhaps any numerical method). The proposed algorithm involves pre-processing the flow variables on a stencil using a conventional shock capturing method and feeding those results into a neural network. The neural network then perturbs the results of the shock capturing method. Post-processing is then applied to the output of the neural network to guarantee consistency \cite{bar2019learning} (or, more generally, could be used to enforce other desirable properties). Hence, the augmented numerical scheme takes on many properties of the original. For example, applying the method to WENO5-JS results in an upwind-biased finite volume method with coefficients that depend on the local solution. The steps of the algorithm for enhancing WENO5-JS can be seen in algorithm \ref{algo}.

\begin{algorithm}[tb]
   \caption{WENO-NN Algorithm}
   \label{alg:train}
\begin{algorithmic}
\STATE  Begin with cell averages $\bar{u}_{j-2:j+2}$\;
\STATE  Scale the cell averages\;
\STATE  Compute coefficients $\tilde{c}_{j-2:j+2}$ with WENO5-JS\;
\STATE  Compute change in coefficients $\Delta \tilde{c}_{j-2:j+2}$ with neural network\;
\STATE  Compute new coefficients $\hat{c}_{j-2:j+2}=\tilde{c}_{j-2:j+2}-\Delta \tilde{c}_{j-2:j+2}$\;
\STATE  Compute final coefficients ${c}_{j-2:j+2}$ by transforming $\hat{c}_{j-2:j+2}$\;
\STATE  Compute cell edge value $u_{j+1/2}={c}_{j-2:j+2} \cdot \bar{u}_{j-2:j+2}$\;
\end{algorithmic}
\end{algorithm}

We use WENO5-JS to pre-process the input data, so that the input to the neural network is the set of finite-volume coefficients found by WENO5-JS. We found that including this pre-processing step significantly improved performance. Once the nonlinear weights $w_i$ are determined according to the WENO5-JS algorithm, the coefficients for each cell average are computed as

\begin{equation}
    \begin{split}
\tilde{c}_{-2} = \frac{1}{3}w_1,\\
\tilde{c}_{-1} = -\frac{7}{6}w_1 - \frac{1}{6}w_2,\\
\tilde{c}_{0} = \frac{11}{6}w_1 + \frac{5}{6}w_2 + \frac{1}{3}w_3,\\
\tilde{c}_{1} = \frac{1}{3}w_2 + \frac{5}{6}w_3,\\
\tilde{c}_{2} = -\frac{1}{6}w_3.
    \end{split}
\end{equation}

These five coefficients are the inputs to the neural network, which outputs a change in each coefficient, $\Delta \tilde{c}_i$. Our neural network uses 3 hidden layers, each with 3 neurons. We deliberately make the network as small as possible to reduce the computational cost of evaluating it. We are able to use such a small network because assuming that the WENO5-JS coefficients are a useful model input is a strong prior, so WENO5-JS performs a significant amount of the required processing. $L_2$ regularization is applied to the output of the neural network to penalize deviations from WENO5-JS, which encourages the network to only change the answer supplied by WENO5-JS when an improved result is expected. The new coefficients are computed by subtracting the change in coefficients from the old coefficients. 

Additionally, the size of the input space is reduced by scaling cell averages within the stencil as \begin{equation}
\bar{\textbf{u}}_s=\frac{\bar{\textbf{u}}-\min{\bar{\textbf{u}}}}{\max{\bar{\textbf{u}}}-\min{\bar{\textbf{u}}}}.
\end{equation}

If all the cell averages have the same value, the scaling equation fails so the value at the cell edge is simply set to the cell average value.

To guarantee that WENO-NN is consistent, we apply an affine transformation to these coefficients that guarantees that they sum to one \cite{bar2019learning}. We derive this transformation by solving the  optimization problem

\begin{equation}
\begin{array}{rrclcl}
\displaystyle \min_{\textbf{c} \in R^5} & \multicolumn{2}{l}{\sum_{n=-2}^{2} (c_n-\hat{c}_n)^2}\\
\textrm{s.t.} & {\sum_{n=-2}^{2} (c_n) = 1},\\
\end{array}
\end{equation}
which can be reformulated with the substitution $\Delta \textbf{c}=\textbf{c}-\hat{\textbf{c}}$ to pose the problem as finding the minimum norm solution to an under-constrained linear system
\begin{equation}
\begin{array}{rrclcl}
\displaystyle \min_{\Delta \textbf{c} \in R^5} & \multicolumn{2}{l}{\sum_{n=-2}^{2} (\Delta c_n)^2}\\
\textrm{s.t.} & {\sum_{n=-2}^{2} (\hat{c}_n+\Delta c_n) = 1},\\
\end{array}
\end{equation}
which has the analytical solution
\begin{equation}
\Delta c_i = \frac{1-\sum_{n=-2}^2 \hat{c}_n}{5}.
\end{equation}

One can use the same approach to enforce arbitrarily high orders of accuracy since the optimization problem has an analytical solution for any constraint matrix of sufficiently high rank

\begin{equation}
\begin{array}{rrclcl}
\displaystyle \min_{\Delta \textbf{c} \in R^5} & \multicolumn{2}{l}{\sum_{n=-2}^{2} (\Delta c_n)^2}\\
\textrm{s.t.} & {A(\hat{\textbf{c}}+\Delta \textbf{c})=\textbf{b}}.\\
\end{array}
\end{equation}

This optimization problem has analytical solution $\Delta \textbf{c}=A^T(AA^T)^{-1}(\textbf{b}-A\hat{\textbf{c}})$ when $AA^T$ is invertible.

We verify that our constraint is satisfied by looking at the convergence rate of WENO-NN for a smooth solution. For this test case, we will simply use WENO-NN and WENO5-JS to take the derivative of $u(x)=\sin(4\pi x) + \cos(4\pi x)$, and compare the results to the analytical solution $\frac{\partial u}{\partial x}^*=-4\pi \sin(4\pi x) + 4\pi \cos(4\pi x)$ using the error metric 

\begin{equation}
E = \sqrt{\frac{||\frac{\partial u}{\partial x}-\frac{\partial u}{\partial x}^*||_2}{N}}.
\end{equation}
 
\begin{figure}[h!]
\centering
\includegraphics[width=0.45\textwidth]{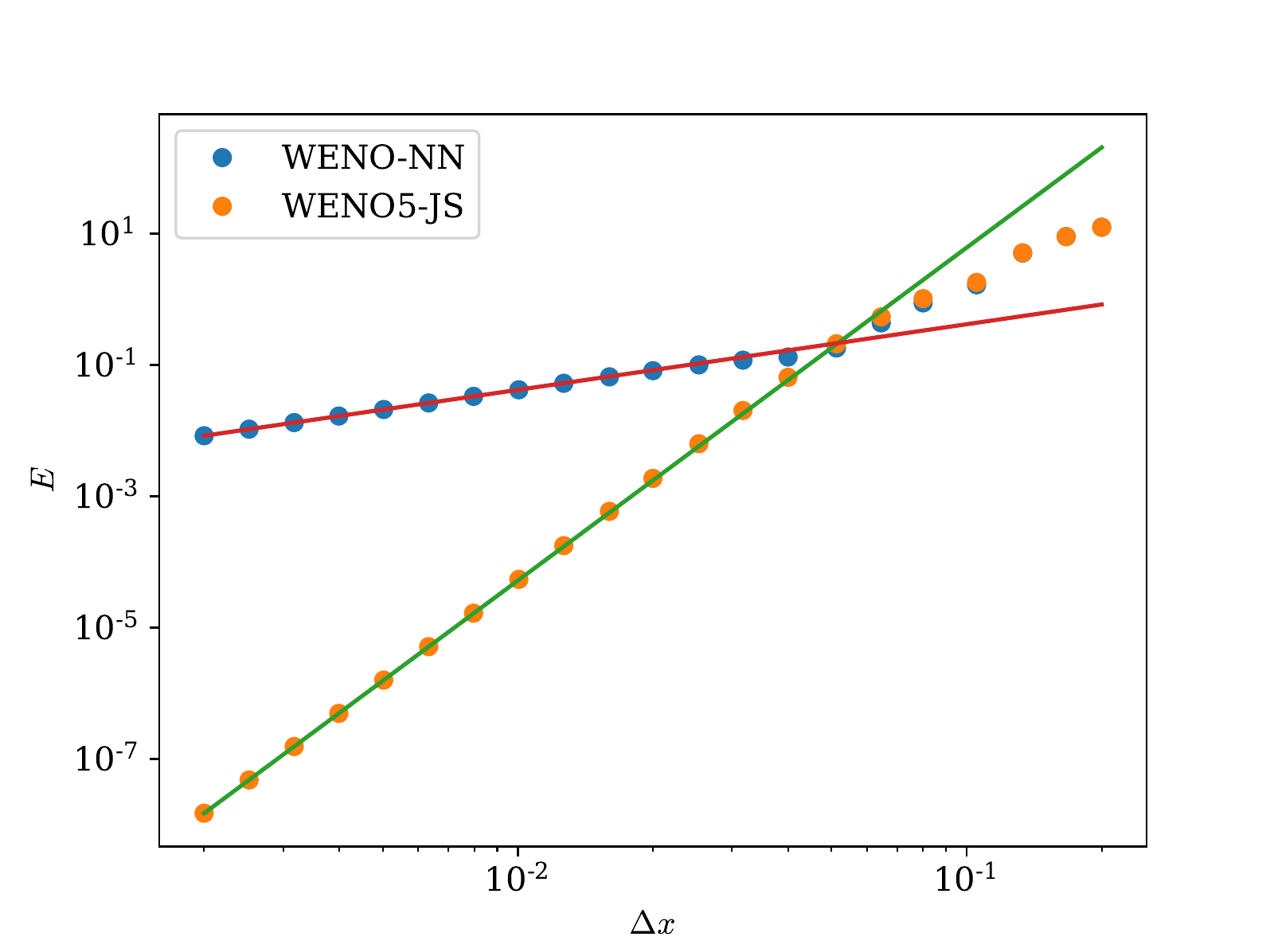}
\caption[smoothConv]{Convergence of WENO-NN and WENO5-JS for smooth solutions}
\label{smoothConv}
\end{figure}

 In Figure~\ref{smoothConv}, we can see that WENO-NN achieves first order accuracy, which confirms that the constraint is satisfied. We also see that, as expected, WENO5-JS converges at fifth order as $\Delta x \rightarrow 0$. However, when discontinuities are present it is not possible to achieve better than first order accuracy with any finite volume method \cite{leveque2002finite}. Despite this fact, it is advantageous to use WENO5-JS over WENO3-JS in such situations, as WENO5-JS still tends to give lower error in discontinuous problems \cite{shu1998essentially}, which is why we chose to use WENO5-JS for processing the cell average values despite the fact that WENO-NN ends up being first-order accurate. Similarly, we see that for some discontinuous problems, WENO-NN gives lower error than WENO5-JS. If a higher order of accuracy is desired in smooth regions of the flow, one could develop a hybrid method using WENO-NN and any high-order method.

\subsection{Other Numerical Methods Used}
For all simulations shown, we use a third-order TVD Runge-Kutta scheme \cite{gottlieb1998total} as our time-stepping method

\begin{equation}
    \begin{split}
u^{(1)} = u^{(n)} + \Delta tL(u^{(n)}),\\
u^{(2)} = \frac{3}{4}u^{(n)} + \frac{1}{4}u^{(1)} + \frac{1}{4}\Delta tL(u^{(1)}),\\
u^{(n+1)} = \frac{1}{3}u^{(n)} + \frac{2}{3}u^{(2)} + \frac{2}{3}\Delta tL(u^{(2)}).
    \end{split}
\end{equation}

For flux-splitting, we use a Lax-Friedrichs flux splitting procedure \cite{shu2003high}

\begin{equation}
\begin{array}{rrclcl}
f^{\pm}(u) = \frac{1}{2}(f(u)\pm \alpha u),\\
\\
\alpha=\displaystyle \max_{u}|f'(u)|.\\
\end{array}
\end{equation}

\begin{figure*}[h!]
\centering
\includegraphics[width=0.9\textwidth]{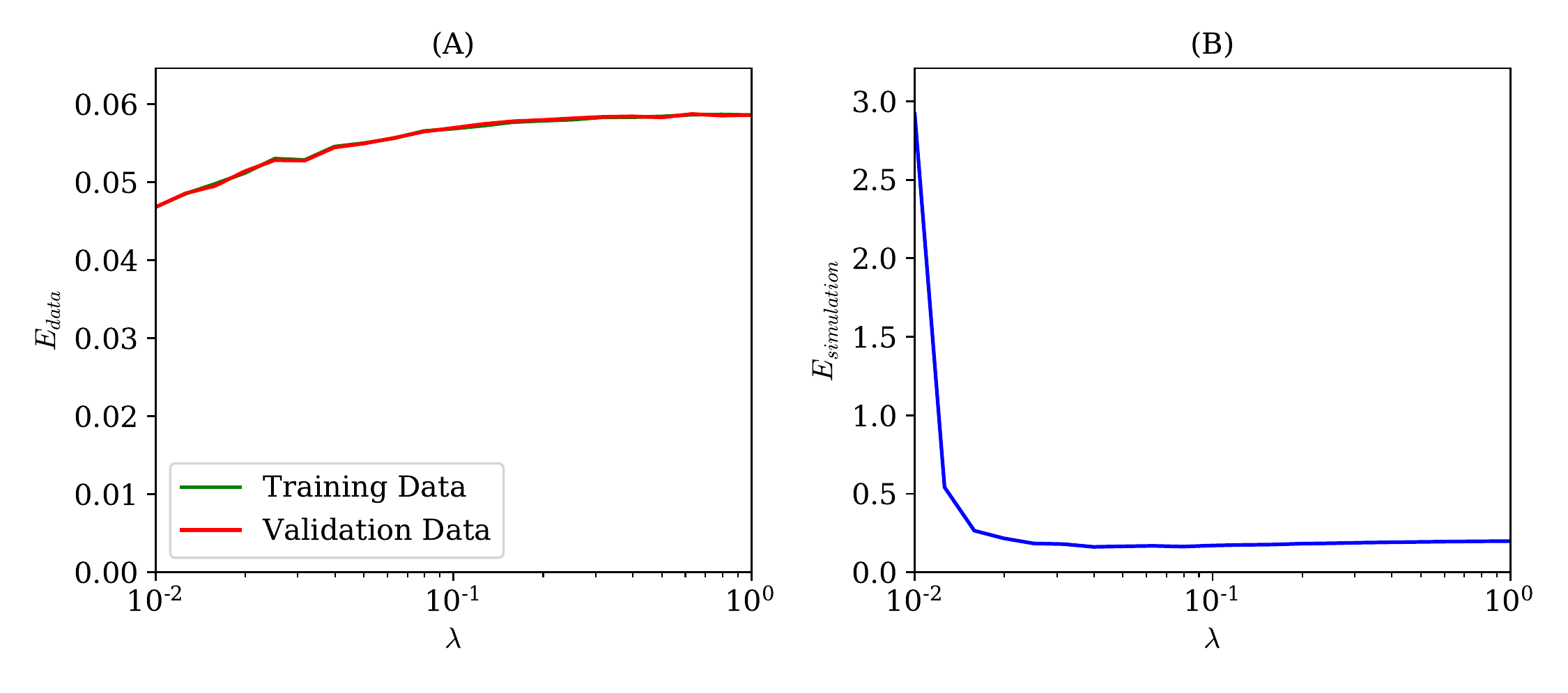}
\caption[RegVErr]{Comparing error trends between (A) exact generated data and (B) simulation results}
\label{RegVErr}
\end{figure*}

In this expression, $f(u)$ is defined as the flux of a 1-D hyperbolic conservation law $\frac{\partial u}{\partial t}+\frac{\partial f(u)}{\partial x}=0$. When solving the 1-D Euler equations, we apply the flux splitting to the characteristic decomposition of the system. For our numerical Riemann solver, we use the Lax-Friedrichs method \cite{chu1979numerical}.

\section{Machine Learning Methodology}
We construct our training data directly from known functions that we expect to represent the waveforms that WENO-NN will encounter in practice. For each datapoint, we start with some function $u(x)$ and a discretized domain of $n$ cells. The cell average is evaluated on each cell as 
\begin{equation}
\bar{u}(x_i)=\frac{1}{\Delta x} \int_{x_i-\frac{\Delta x}{2}}^{x_i+\frac{\Delta x}{2}}u(x) dx,
\end{equation}

and because we chose the form of $u(x)$ we can evaluate the cell average analytically. We also evaluate the function value on the cell boundary as $u(x_i+\Delta x/2)$ analytically. We then move along the domain and form the dataset based on the stencil size. So for WENO-NN, one datapoint involves 5 cell averages as the input with the function value on the cell boundary as the output. The functions we use when creating the dataset are step functions, sawtooth waves, hyperbolic-tangent functions, sinusoids, polynomials, and sums of the above.  

When adding a new entry to the dataset, we first check to see if it is close to other points already present in the $L_2$ sense.  Sufficiently close points are not added to the dataset to prevent redundant data that will slow down the training process. The resulting dataset has 75241 entries. 

When training the network, we use the Adam optimizer \cite{kingma2014adam}, split the data into batches of 80 points to estimate the gradient, and optimize for 10 epochs using the Keras package in python \cite{chollet2015keras}. We trained the network from many different randomly chosen initial guesses of the parameters, and chose the best one based on performance in simulating the linear advection of a step function. We apply $L_2$ regularization with a constant of $\lambda=0.1$ to the neural network output, and find that when splitting the data into a training set of 80\% of the data and a validation set of the other 20\% of the data our in-sample error is 0.569 and the out-of-sample error is 0.571, averaged from 100 trials of training on the dataset, so overfitting within the generated dataset is not a concern. This difference is so small because the model we are training is of relatively low complexity, and is essentially underfitting the generated dataset.  We use mean squared loss as our objective function to minimize.

Despite the fact that we do not see overfitting within the generated dataset, we still observe overfitting when we apply the method to an actual simulation. Figure~\ref{RegVErr} shows the average training error, average validation error, and average error when using the method to simulate a PDE for different regularization values $\lambda$ of the neural network output. The training and validation error are computed using the mean square error,

\begin{equation}
E_{data} = \frac{\sum_{i=1}^N(y_i-y_i^*)^2}{N},
\end{equation}

while the simulation error is computed by using the learned numerical method to linearly advect a step function and computing the $L_2$ error at the end of the simulation,

\begin{equation}
E_{simulation} = \sqrt{\int_0^L(\bar{u}(x,T)-\bar{u}^*(x,T))^2dx}.
\end{equation}

One can see that adding regularization causes error to increase in both the training and validation datasets but decreases the error in the simulation results. Hence, we can see that we are overfitting to the training data, but because the validation data does not show this, we can conclude that the dataset does not exactly match the distribution we are trying to approximate. 

\section{Results}
\subsection{Advection Equation}
These results will focus on comparing WENO5-JS to WENO-NN. Every WENO-NN result we show in this paper was generated using the same neural network with the same weights. As such, our numerical method is broadly applicable to problems not discussed in this paper, in contrast with many machine learning solutions that are problem-specific. The first test case we look at is the linear advection of a step function on a periodic domain. Mathematically, this IBVP is posed as

\begin{equation}
    \begin{split}
\frac{\partial u}{\partial t} + c\frac{\partial u}{\partial x} = 0,\\
    u(0,x)= 
\begin{cases}
    1,& \text{if } x\geq L/2,\\
    0,              & \text{otherwise,}
\end{cases}\\
u(t,0)=u(t,L).
    \end{split}
\end{equation}

\begin{figure*}[h!]
\centering
\includegraphics[width=0.9\textwidth]{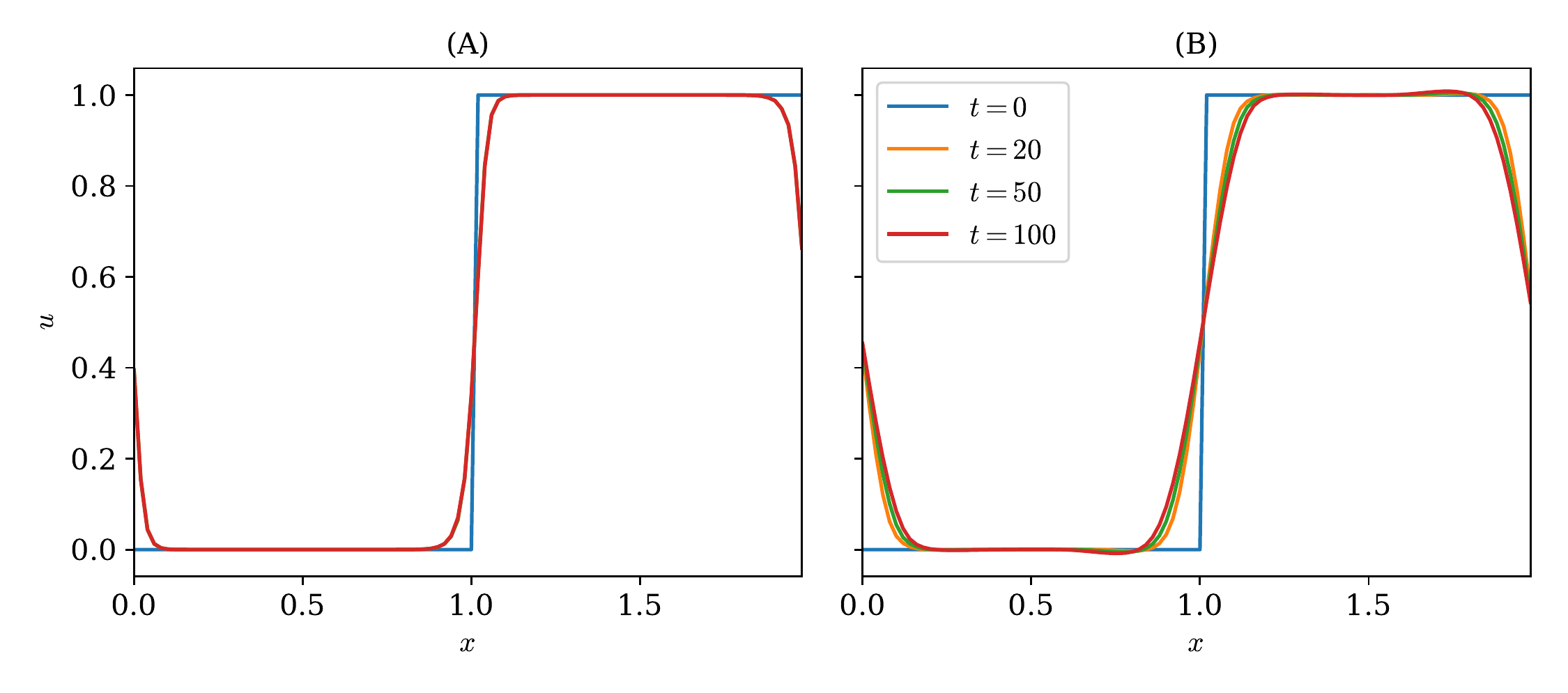}
\caption[AdvSol]{Numerical solutions of the advection equation at $t=0, 20, 50$ and $100$ using (A) WENO-NN and (B) WENO5-JS. Note that the curves in (A) for $t>0$ are indistinguishable.}
\label{AdvSol}
\end{figure*}

For this simulation, we set $c=1$ and $L=2$. We split the domain into 100 cells, use a CFL number of 2/3, and run the simulation for 50 periods for a total time of $T=100$. Figure~\ref{AdvSol} shows the solution of this PDE for WENO5-JS and WENO-NN at $t=0,20,50$ and $100$. The solution at $t=0$ is also the exact solution at all the other times plotted.

One can see that the solution using WENO-NN provides a closer visual fit to the exact solution, as WENO5-JS diffuses the discontinuity more significantly than WENO-NN. WENO5-JS also introduces noticeable overshoot behind the discontinuity. The neural network has the interesting property that the waveform is nearly invariant to its propagation, while WENO5-JS continues to diffuse the solution. This behavior can be explained by examining the artificial fluid properties associated with the modified equation obtained by Taylor series expansion (assuming linearity of the scheme). The modified PDE is
\begin{equation}
\frac{\partial u}{\partial t} + c\frac{\partial u}{\partial x} = \nu\frac{\partial^2 u}{\partial x^2} + \delta\frac{\partial^3 u}{\partial x^3} - \sigma\frac{\partial^4 u}{\partial x^4} + \ldots
\end{equation}
The expansions give expressions for the artificial viscosity, dispersion, and hyperviscosity, $\frac{\partial\bar{u}}{\partial t}+\frac{u(x+\frac{\Delta x}{2})-u(x-\frac{\Delta x)}{2}}{\Delta x}=0$ after making the substitutions $u(x+\frac{\Delta x}{2})=\sum_{n=-2}^2c_n\bar{u}(x+n\Delta x)$ and $u(x-\frac{\Delta x}{2})=\sum_{n=-2}^2c_n\bar{u}(x+(n-1)\Delta x)$, and are computed as

\begin{eqnarray}
\nu & = \Delta x \sum_{n=-2}^{2}c_n\frac{(n-1)^2-n^2}{2}, \\
\delta & = \Delta x^2 \sum_{n=-2}^{2}c_n\frac{(n-1)^3-n^3}{6}, \\
\sigma & = -\Delta x^3 \sum_{n=-2}^{2}c_n\frac{(n-1)^4-n^4}{24}.
\end{eqnarray}

\begin{figure*}[h!]
\centering
\includegraphics[width=1\textwidth]{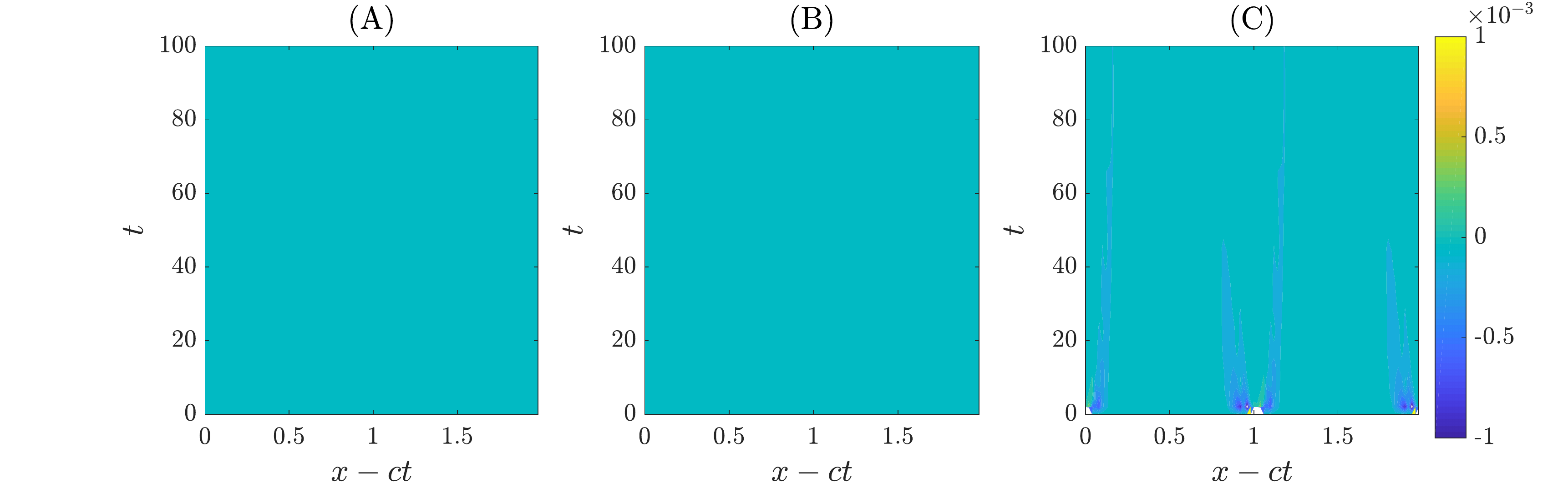}
\caption[JSProp]{Influence of (A) artificial viscosity, (B) dispersion, and (C) hyperviscosity of WENO5-JS}
\label{JSProp}
\end{figure*}

Figure~\ref{JSProp} shows these quantities for WENO5-JS. In order to estimate the contribution of each term, we approximated the higher-order spatial derivatives using standard finite-volume methods, and scale each by the magnitude of that derivative. For example, the influence of artificial viscosity is computed as

\begin{eqnarray}
\bar{\nu}(x)=\frac{\nu(x+\Delta x/2)+\nu(x-\Delta x/2)}{2}, \\
|u''(x)|=|{\frac{u'(x+\Delta x/2)-u'(x-\Delta x/2)}{\Delta x}}|, \\
I_{\nu}(x)=\bar{\nu}(x)|u''(x)|
\end{eqnarray}

Hence, we ignore regions of the flow where the coefficient may signify that artificial viscosity (or other properties) are being added when they would have a negligible effect because the derivative is small.

One can see that for WENO-JS there is no viscosity or dispersion, as the method is designed such that on each substencil $\sum_{n=-2}^{2}c_n\frac{(n-1)^2-n^2}{2}=0$ and  $\sum_{n=-2}^{2}c_n\frac{(n-1)^3-n^3}{6}=0$, so WENO5-JS applies only hyperviscosity. The method  applies a small amount of negative hyperviscosity near the discontinuity. As time goes on and the discontinuity continues to diffuse, the influence of hyperviscosity decreases. 

\begin{figure*}[h!]
\centering
\includegraphics[width=1\textwidth]{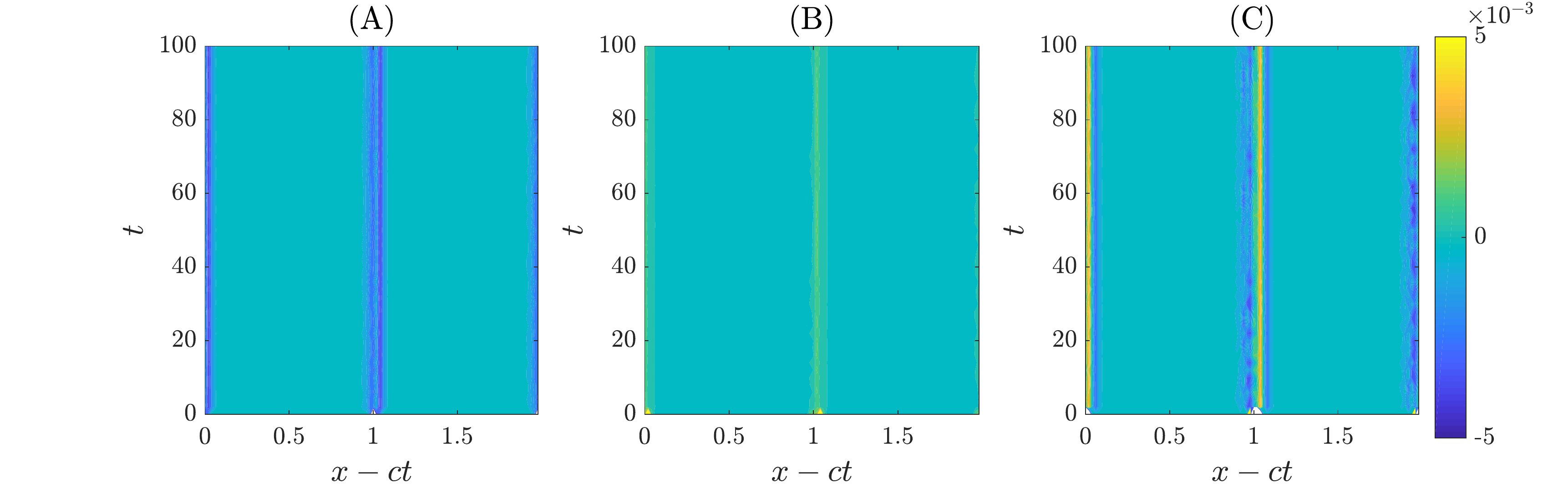}
\caption[NNProp]{Influence of (A) artificial viscosity, (B) dispersion, and (C) hyperviscosity of WENO-NN}
\label{NNProp}
\end{figure*}

Figure \ref{NNProp} shows that unlike WENO5-JS, WENO-NN adds both artificial viscosity and dispersion to the solution. We see that near the discontinuity, negative viscosity is being added, which apparently provides the anti-diffusion that causes the discontinuity to retain its steepness, while hyperviscosity is applied to stabilize the solution.

\begin{figure*}[h!]
\centering
\includegraphics[width=1\textwidth]{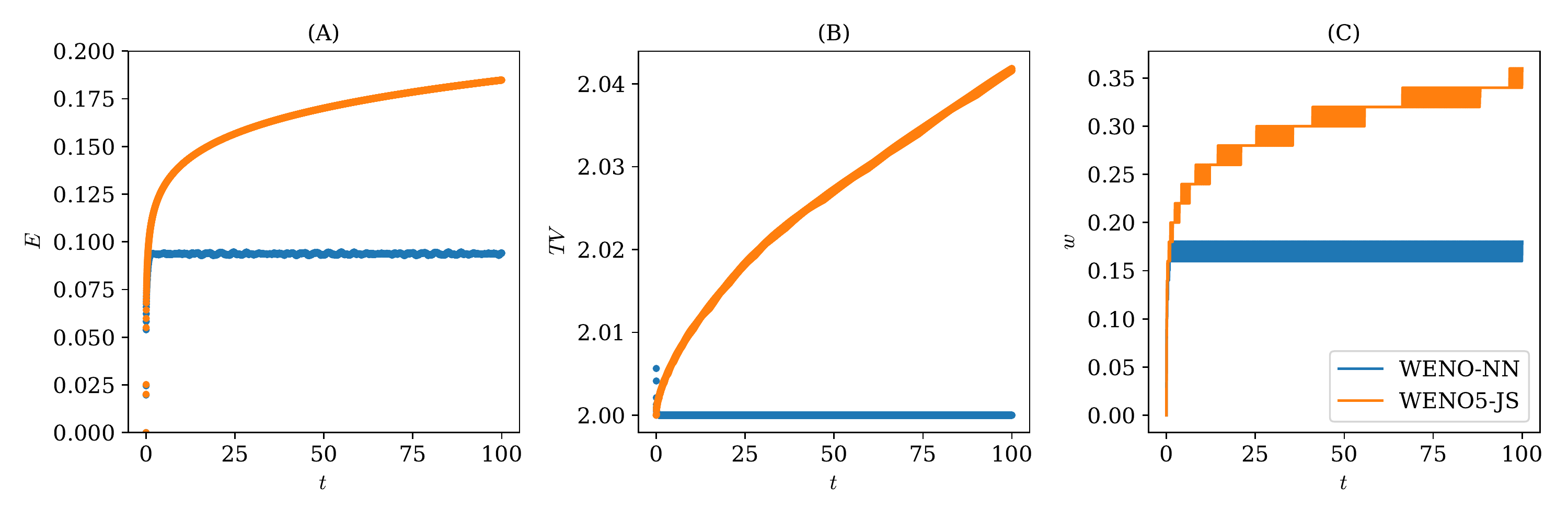}
\caption[AdvErrs]{Comparing (A) $L_2$ error, (B) total variation, and (C) discontinuity width over time for WENO-NN and WENO-JS}
\label{AdvErrs}
\end{figure*}

We obtain a quantitative picture of the error in figure~\ref{AdvErrs}.  We plot the $L_2$ error over time (measured to the exact solution), as well as the total variation, $TV = \sum_{i=1}^N|u(\Delta x i)-u(\Delta x (i-1))|$, to indicate when oscillations have been induced in the solution.  We also measured the width over which the discontinuity is spread by counting the cells that have an error above a certain threshold (in this case chosen to be 0.01) and multiplying this number by $\Delta x/2$ since there are two discontinuities in the simulation. 

\begin{figure*}[h!]
\centering
\includegraphics[width=0.9\textwidth]{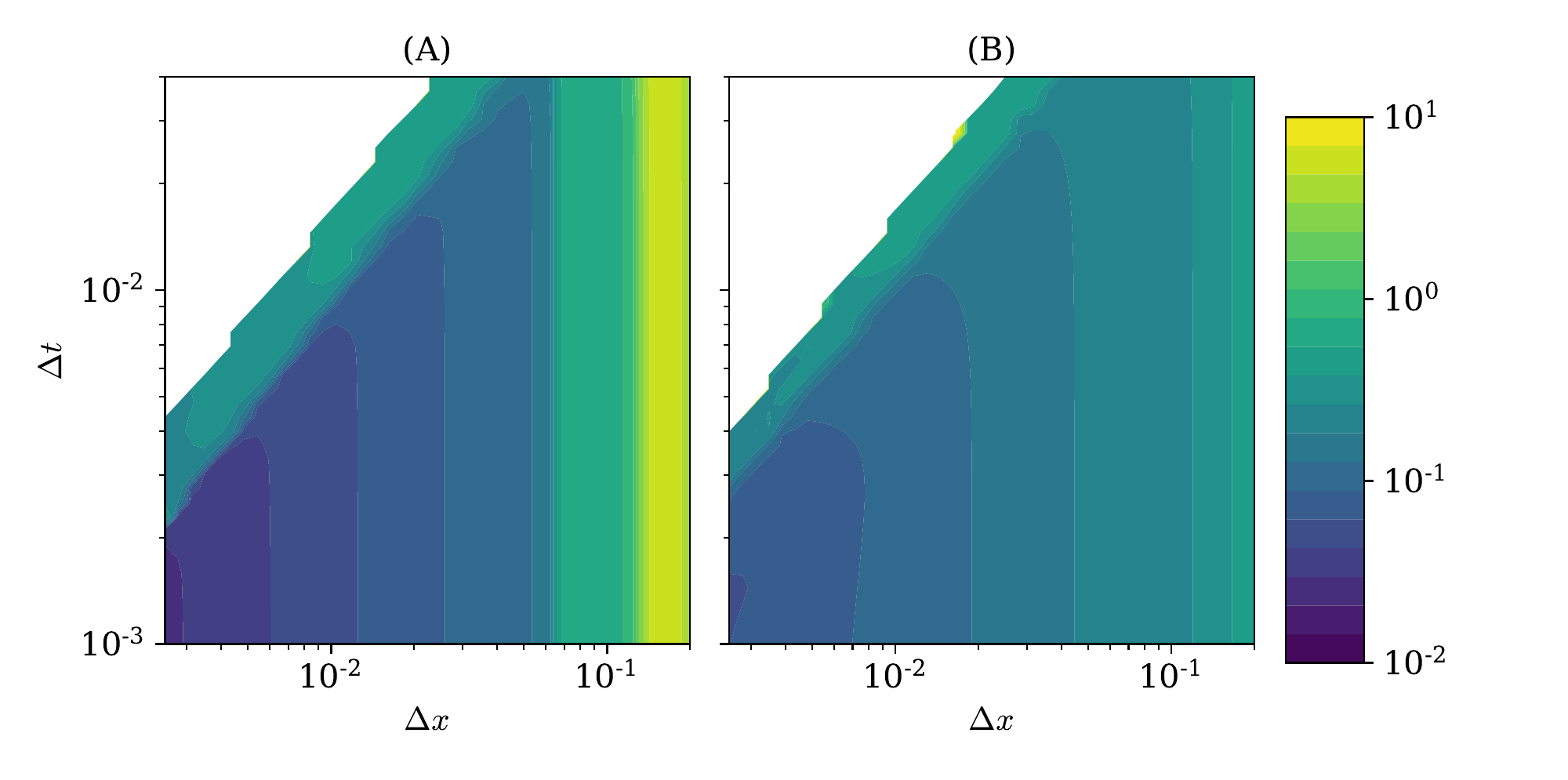}
\caption[L2Errs]{$L_2$ error at the end of the simulation for (A) WENO-NN and (B) WENO5-JS}
\label{L2Errs}
\end{figure*}

The figure shows that WENO-NN decreases the error by almost a factor of 2 \footnote{Note that the error oscillates between two different values because in the exact solution the discontinuity switches between being on the edge of a cell and 1/3 of a cell width away from either the left or right of a cell edge since the CFL number is 2/3. To get a smooth curve, we apply a filter to the error and plot $E(i) = \frac{e(i)+e(i-1)+e(i-2)}{3}$}. Although the total variation spikes at the start of the WENO-NN simulation, it is damped out and returns back to approximately the true value of 2, while the WENO5-JS total variation steadily climbs to above 2.04. We see a similar behavior in the discontinuity width, where WENO-NN reaches its steady value relatively quickly, while WENO5-JS continues to spread.

In order to determine how WENO-NN performs in different settings, the spatial and temporal discretizations were varied, and the $L_2$ error at the end of the simulation was measured.  We again use a domain of length 2 and simulate for 50 periods. These results can be seen in figure~\ref{L2Errs}.

\begin{figure}[h!]
\centering
\includegraphics[width=0.45\textwidth]{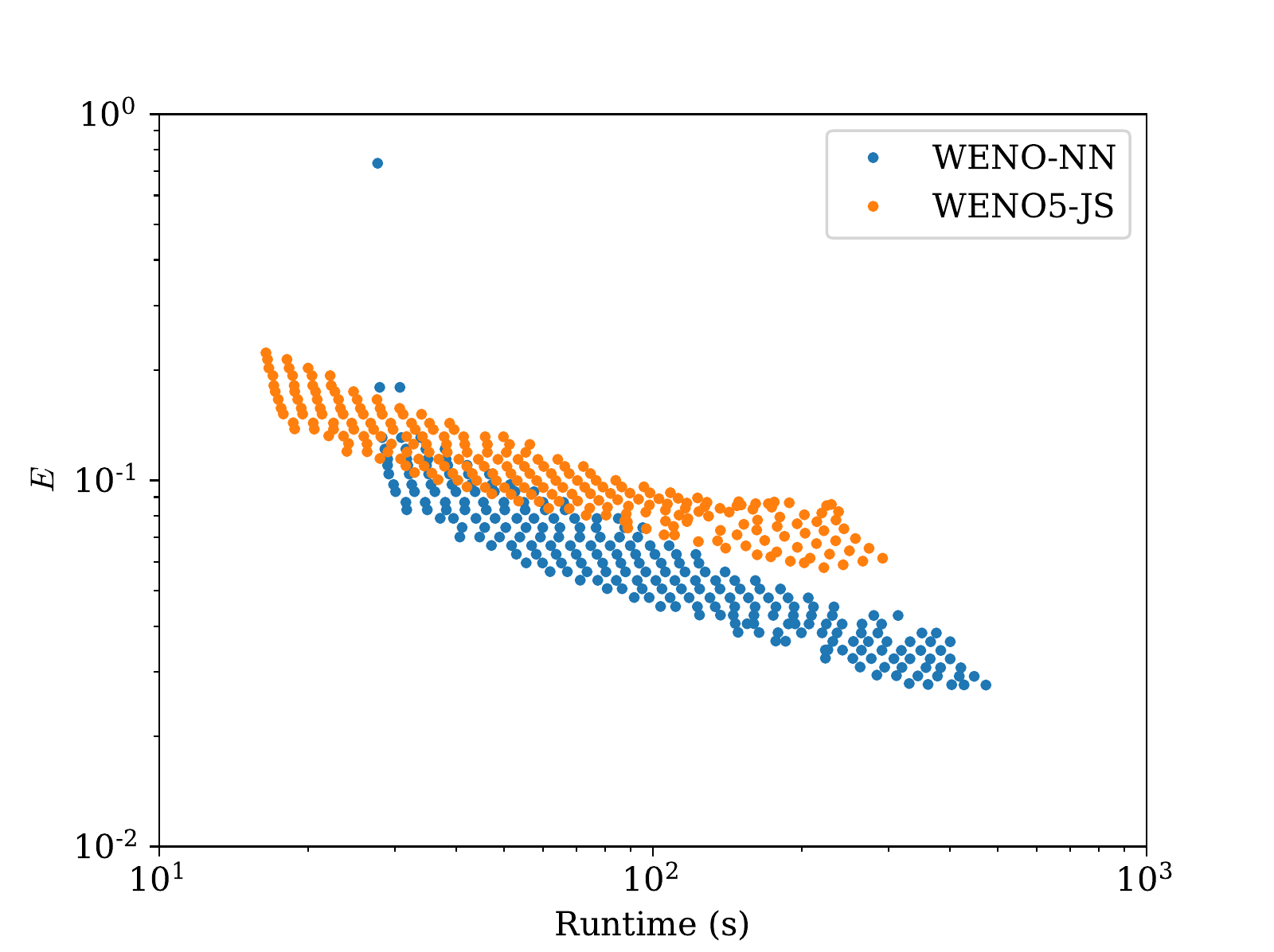}
\caption[RTvErr]{Comparing the $L_2$ error and runtime of WENO-NN and WENO5-JS for $0.25<\text{CFL}<0.75$ and $\Delta x < 0.025$}
\label{RTvErr}
\end{figure}

We can see that WENO-NN tends to outperform WENO5-JS in regions where the spatial discretization is fine, but results in a larger $L_2$ error for coarse discretizations. To further compare the methods, figure~\ref{RTvErr} shows the error against the runtime for the two methods within a range of CFL values. We will only look at moderate CFL numbers, between 0.25 and 0.75, as stability becomes a concern for both methods above this range, and it becomes inefficient to run the simulation with CFL numbers below this range. We will also restrict the cell width to be below 0.025, as coarser meshes cause the final waveform to be unrecognizable compared to the exact solution for both methods, so the error comparison becomes meaningless. We see that when the CFL number is of a moderate value and the grid is sufficiently refined, WENO-NN typically achieves lower errors with smaller run time than WENO5-JS.

\subsection{Inviscid Burgers' Equation}

We next consider the inviscid Burgers' equation. Unlike the linear advection equation that included only contact (initial) discontinuities, the inviscid Burgers' equation results in shocks from smooth initial data. The distinction here is important: for a shock, the dynamics of the PDE will drive the solution towards a discontinuity, countering any diffusive effects associated with the numerics. We will again consider periodic boundary conditions, though we will start the simulation with a Gaussian as the initial condition. Hence, the IBVP is posed as

\begin{eqnarray}
\frac{\partial u}{\partial t} + \frac{1}{2}\frac{\partial u^2}{\partial x} = 0, \\
u(0,x)=\exp(-k(x-\frac{L}{2})^2), \\
u(t,0)=u(t,L).
\end{eqnarray}

\begin{figure}[h!]
\centering
\includegraphics[width=0.45\textwidth]{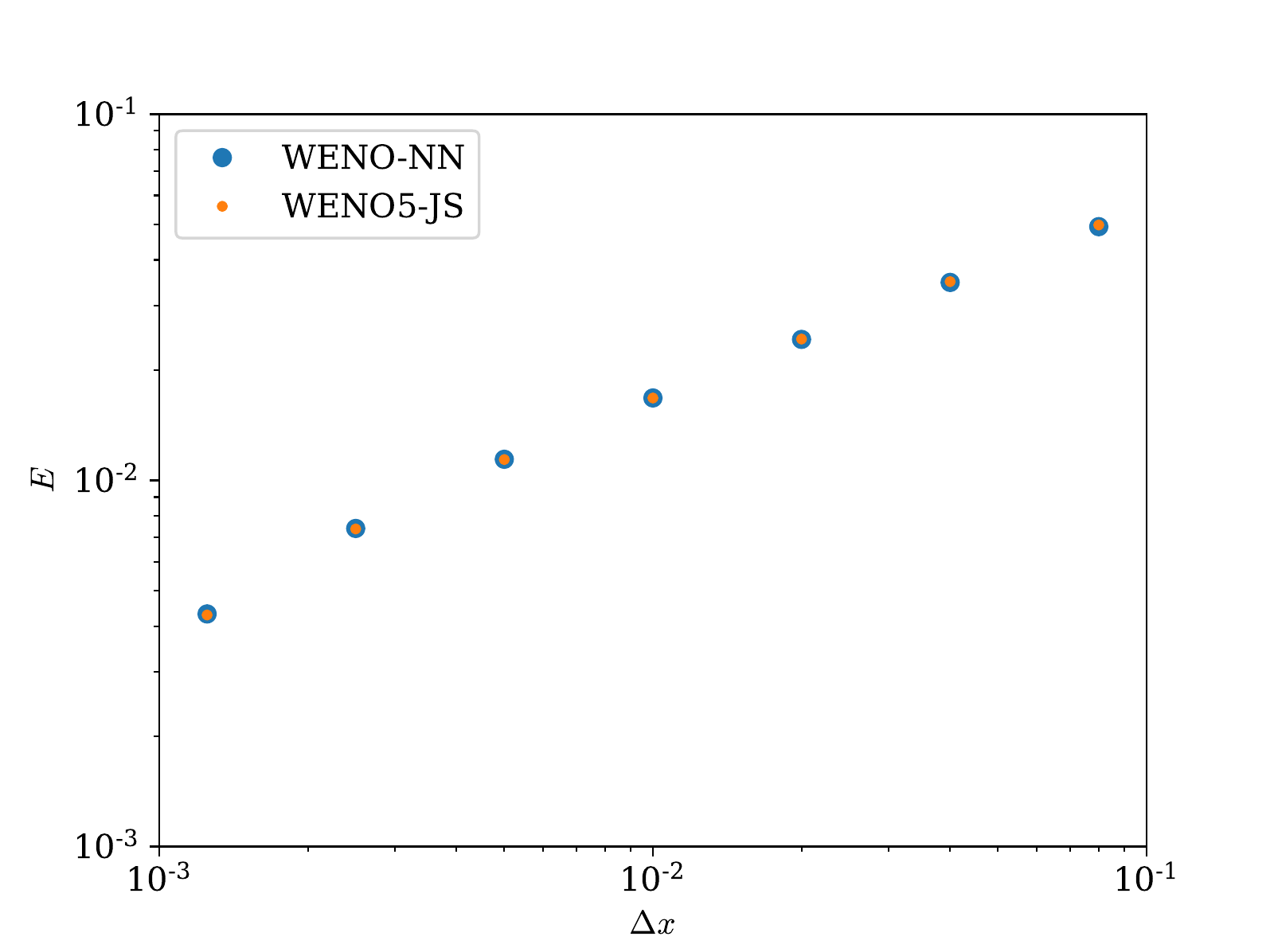}
\caption[InvConv]{Comparing error vs. grid spacing of WENO-NN and WENO5-JS for the inviscid Burgers' equation}
\label{InvConv}
\end{figure}

We simulate the problem for a time of $T=4$ on a domain of length $L=2$, and a value of $k=20$. We first approximate the exact solution by solving this simulation with $\Delta x=3.125\cdot 10^{-4}$ and $\Delta t=1.5625\cdot 10^{-4}$ for a total of 6400 cells and 25601 timesteps. What we see is that the $L_2$ error is roughly the same for WENO5-JS and WENO-NN, as shown by Figure~\ref{InvConv}. Hence, we should expect the method to perform similarly to WENO5 in the presence of a shock.

\subsection{1-D Euler Equations}

The last test case we will look at is the Shu-Osher problem, a test case involving the 1-D Euler equations. Note that the method was also tested on the Sod problem, but because this test case did not lead to any conclusions not drawn from either the advection equation or the inviscid Burgers' equation, these results have been omitted. The Shu-Osher problem is a model problem for turbulence-shockwave interactions. It involves the following equations and initial conditions

\begin{eqnarray}
    \frac{\partial \rho}{\partial t} + \frac{\partial (\rho u)}{\partial x}=0, \\
    \frac{\partial \rho u}{\partial t} + \frac{\partial (P + \rho u^2)}{\partial x}=0, \\
    \frac{\partial E}{\partial t} + \frac{\partial ((E+P)u)}{\partial x}=0, \\
    P = (\gamma-1)(E-\frac{1}{2}\rho u^2), \\
\rho(0,x)= 
\begin{cases}
    3.857143,& \text{if } x\leq 1\\
    1+\epsilon\sin(5x),              & \text{otherwise}
\end{cases},\\
u(0,x)= 
\begin{cases}
    2.629369,& \text{if } x\leq 1\\
    0,              & \text{otherwise}
\end{cases},\\
P(0,x)= 
\begin{cases}
    10.33333,& \text{if } x\leq 1\\
    1,              & \text{otherwise}
\end{cases}.
\end{eqnarray}

\begin{figure*}[h!]
\centering
\includegraphics[width=1.0\textwidth]{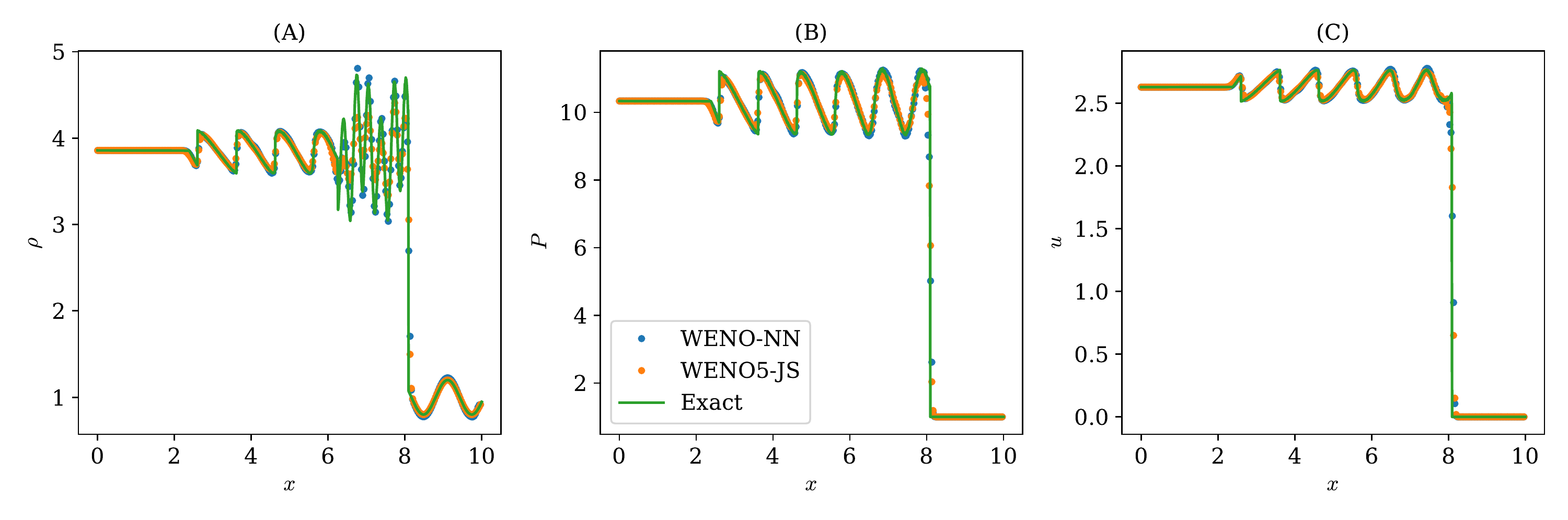}
\caption[ShuSol]{Comparing (A) density, (B) pressure, and (C) velocity of WENO-NN and WENO5-JS to the exact solution for the Shu-Osher problem}
\label{ShuSol}
\end{figure*}

\begin{figure*}[h!]
\centering
\includegraphics[width=0.9\textwidth]{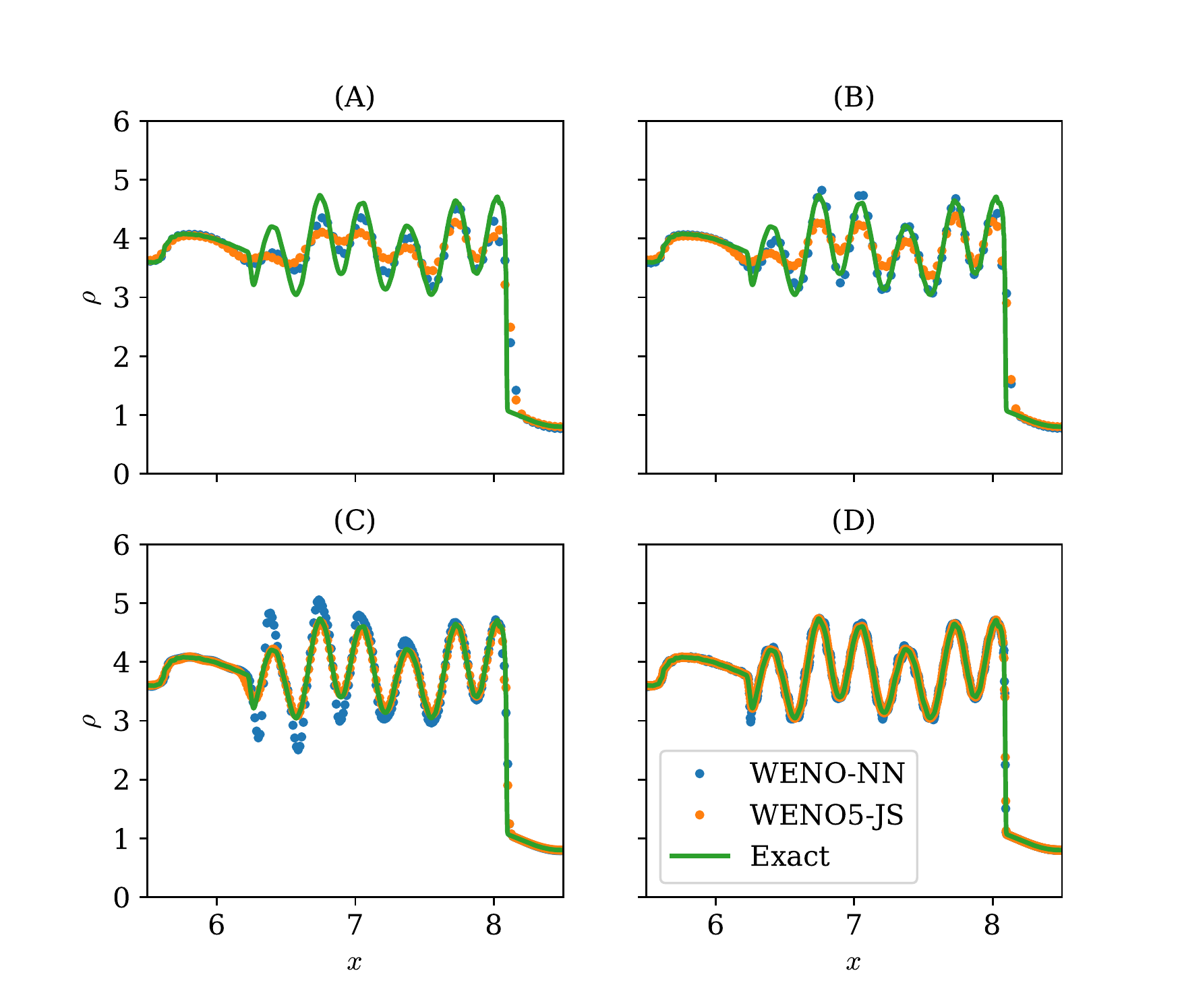}
\caption[ShuConv]{Zoomed in view of the turbulent section for different grid resolutions of (A) 250, (B) 300, (C) 800, and (D) 3200 cells for the Shu-Osher problem}
\label{ShuConv}
\end{figure*}

The simulation takes place on a domain of length $L=10$ and is run until a final time of $T=2$. $\epsilon$ is set to 0.2. We first obtain an approximately exact solution by discretizing the solution into 12800 cells and 10240 time-steps and use WENO5-JS for the simulation. This grid is fine enough to consider the solution exact. We then solve the problem using 300 cells and 240 time-steps using both WENO5-JS and WENO-NN, and compare the numerical results to the exact solution. Figure~\ref{ShuSol} shows the density, pressure, and velocity at the end of the simulation.

The most interesting aspect of the solution is the highly oscillatory section of the density profile, which is considered to behave similarly to turbulence. Figure~\ref{ShuConv} shows a zoomed in view of this section at different grid resolutions.

One can see that the neural network diffuses the oscillations significantly less than WENO5 for coarse grids, which is an encouraging result in terms of simulating actual turbulence. As the mesh is further refined, the WENO-NN appears to overcompensate, though on the very fine grid both WENO-J5 and WENO-NN are similar (provided WENO-NN is stable, then it is constrained to converge as at least first-order).

\section{Discussion and Conclusions}

By training a neural network to process the outputs of the WENO5-JS algorithm, we were able to improve its accuracy, particularly in problems where the artificial diffusion introduced in WENO5-JS was excessive. While WENO-NN is more expensive per evaluation than WENO5-JS, it achieved lower errors on coarser grids, which indicates some potential to be useful more generally. We trace these performance improvements to increased flexibility in the neural network compared to WENO5-JS, as it can add artificial viscosity and dispersion while WENO5-JS coefficients are constrained to make these quantities zero. By analyzing the advection of a step function, we found that WENO-NN applies negative artificial viscosity near the discontinuity, which allows it to maintain its sharp profile (this takes place sometime into the simulation after the initial profile has been slightly smoothed due to artificial viscosity that prevents spurious oscillations). We then observe similar behavior in the Shu-Osher problem, where we see that WENO5-JS diffuses the fine features of the solution more than WENO-NN. However, we also found that at certain resolutions WENO-NN applies too much negative artificial viscosity, resulting in too much amplification of these fine scale features, though this amplification does not develop into an instability. For true shocks, as opposed to contact discontinuities, we found that our method performs very similarly to WENO5-JS. 

There is still room for improvement for the method. We believe that the development of training data could be improved. Perhaps better results could be achieved by formulating the training data in a more systematic way. Some papers generate the data directly from simulations \cite{bar2019learning}. While this approach results in neural networks that are specific to the equation that they were trained on, it could lead to methods that outperform methods designed to work well for general problems. Another drawback of WENO-NN is that it does not inherit the high-order convergence of WENO5-JS. It would be an improvement to the method to be able to structure the network such that its coefficients more quickly converge to those of either WENO5-JS or the constant coefficient scheme that maximizes order of accuracy in the presence of smooth solutions. However, this must be done in a way that does not interfere with predictions in non-smooth regimes that benefit from low-order behavior, which is a non-trivial task. Until such a method is developed, one would need to use WENO-NN as part of a hybrid scheme if higher order convergence is desired in smooth regions.

Another outstanding issue with machine-learned schemes is stability.  The WENO-NN scheme used here seemed to inherit the stability of the underlying WENO5-JS scheme that it was based on, but this needn't have been the case, and we cannot offer proof or an estimate for the maximal CFL.

In future work, we aim to test the method on large-scale, multidimensional problems, since we would expect the benefits seen in 1-D problems to be more significant when multiple spatial dimensions are present.


%
\section*{Conflict of interest}
The authors declare that they have no conflict of interest.

\section*{Acknowledgements}
This material is based upon work supported by the National Science Foundation Graduate Research Fellowship under Grant No. 1745301
\bibliography{example_paper}
\bibliographystyle{icml2020}

\end{document}